\documentclass[preprint,11pt]{elsarticle}

\journal{npj Computational Materials}

\makeatletter
\def\ps@pprintTitle{%
    \let\@oddhead\@empty
    \let\@evenhead\@empty
    \let\@oddfoot\@empty
    \let\@evenfoot\@empty}
\makeatother

\usepackage{hyperref}
\hypersetup{backref,colorlinks=true,allcolors=MidnightBlue}
\usepackage{xr}
\usepackage[margin=2.5cm]{geometry}

\usepackage{graphicx}
\graphicspath{{./figures/}}
\usepackage[dvipsnames]{xcolor}
\usepackage{tcolorbox}
\usepackage{amsmath,amssymb}
\usepackage{float}
\usepackage{multirow}
\usepackage{tabularx}
\usepackage{pifont}
\usepackage{subcaption}
\usepackage{wrapfig}
\usepackage{miller}
\usepackage{mathrsfs}
\usepackage{orcidlink}
\usepackage{pdfpages}

\usepackage{tikz}
\usetikzlibrary{calc,arrows,shapes,positioning}
\usetikzlibrary{patterns,snakes}
\usetikzlibrary{positioning,decorations.pathreplacing}

\usepackage{epstopdf}
\AppendGraphicsExtensions{.tif}

\definecolor{sangria}{rgb}{0.57, 0.0, 0.04}
\definecolor{arsenic}{rgb}{0.23, 0.27, 0.29}
\definecolor{prussianblue}{rgb}{0.0, 0.19, 0.33}
\definecolor{phthalogreen}{rgb}{0.07, 0.21, 0.14}
\definecolor{dgreen}{rgb}{0.0, 0.4, 0.2}

\def\fig#1{Fig.~\ref{fig:#1}}
\def\eq#1{Eq.~\eqref{eq:#1}}
\def\tab#1{Table~\ref{tab:#1}}

\begin{document}
\begin{frontmatter}
\title{
AtomProNet: Data flow to and from machine learning interatomic potentials in materials science}
\author[label1]{Musanna Galib \orcidlink{0000-0002-1735-7546}}
\affiliation[label1]{organization={Department of Mechanical Engineering, The University of British Columbia},
            addressline={2054 - 6250 Applied Science Lane}, 
            city={Vancouver},
            postcode={V6T 1Z4}, 
            state={BC},
            country={Canada}
            }
\ead{galibubc@student.ubc.ca}
\author[label1]{Mewael Isiet \orcidlink{0000-0002-0924-2159}}
\author[label1]{Mauricio Ponga \orcidlink{0000-0001-5058-1454}}
\ead{mponga@mech.ubc.ca}
%
\begin{abstract}
As the atomistic simulations of materials science move from traditional potentials to machine learning interatomic potential (MLIP), the field is entering the second phase focused on discovering and explaining new material phenomena. 
%
%
While MLIP development relies on curated data and flexible datasets from ab-initio simulations, transitioning seamlessly between ab-initio workflows and MLIP frameworks remains challenging.
%
A global survey was conducted to understand the current standing (progress and bottleneck) of the machine learning-guided materials science research. 
The survey responses have been implemented to design an open-source software to reduce the access barriers of MLIP models for the global scientific community. 
Here, we present \texttt{AtomProNet}, an open-source Python package that automates obtaining atomic structures, prepares and submits ab-initio jobs, and efficiently collects batch-processed data for streamlined neural network (NN) training. 
Finally, we compared empirical and start-of-the-art machine learning potential, showing the practicality of using MLIPs based on computational time and resources.
\end{abstract}
%

\end{frontmatter}
\section{\label{sec:Introduction}Introduction}

The “holy grail” for computational materials scientists is arguably an exact potential energy surface (PES) representation that underpins the accurate quantum mechanics (QM) data without any approximations applied to the Schrödinger equation~\cite{10.1080/00268976.2017.1333644}.
However, for a system $>10^2$ atoms,  lower-dimensional representations for PES are used to tackle the accuracy-efficiency trade-off~\cite{10.1021/acs.jpca.2c06778}. 
Fortunately, the dissuading requirements for data flow from QM to PES can be avoided by using MLIPs. 
Beginning in 2000s, the scientific community started building databases leveraging the advanced high-throughput computational methods (Materials Project~\cite{10.1063/1.4812323}, the Open Crystallographic Database~\cite{10.1107/S0021889809016690}, AFLOWLIB~\cite{10.1016/j.commatsci.2012.02.002}, the NOMAD archive~\cite{10.1557/mrs.2018.208}, JARVIS~\cite{10.1038/s41524-020-00440-1}, the Materials Commons~\cite{10.1007/s11837-016-1998-7}, Alexandria~\cite{10.1016/j.mtphys.2024.101560}, the Open Quantum Materials Database~\cite{10.1038/npjcompumats.2015.10}, Open Materials 2024 (OMat24)~\cite{arxiv.org/abs/2410.12771}) and high-throughput experimental methods (Open High Throughput Experimental Materials Database~\cite{10.1038/sdata.2018.53}, the Inorganic Crystal Structure Database~\cite{10.1080/08893110410001664882}, the Cambridge Structural Database~\cite{10.1107/S2052520616003954}).
Therefore, the database such as 
Materials Project~\cite{10.1063/1.4812323}, NIST~\cite{10.1038/s41467-019-13297-w} and AFLOWLIB~\cite{10.1016/j.commatsci.2012.02.002}, become game-changer by providing extensive training datasents for MLIPs. 
These high-throughout computational databases are built mostly employing density functional theory (DFT)~\cite{10.1103/PhysRev.136.B864,10.1103/PhysRev.140.A1133} based QM method to mimic deep physical underpinnings from electronic structures. 
These data flows to MLIPs to build the 3-dimensional PES of the configurational space, optimizing the regression parameters to obtain a smooth PES that best interpolates between the reference energies~\cite{10.1016/j.actamat.2021.116980}.
This high dimensionality of the parameter space for the complex optimization problem makes the MLIPs more capable compared to traditional interatomic potentials (IP), as traditional IPs target only a certain set of material properties for training. 
Thus, state-of-the-art MLIPs achieve almost an order of magnitude better prediction accuracy ($\mathrm{<5\: meV/atom}$ in energies and $\mathrm{<0.1\: eV/Å}$ in forces) than traditional IPs; enough resolutions to solve small energy differences between polymorphs~\cite{10.1021/acs.jpca.9b08723}.
\par To build the DFT dataset, a few of the most popular software is the Vienna Ab Initio Simulation Package (VASP)~\cite{10.1016/0927-0256(96)00008-0}, and Quantum Espresso~\cite{10.1063/5.0005082} whereas one of the popular interfaces to implement MLIP into classical molecular dynamics (MD) is the Large-scale Atomic/Molecular Massively Parallel Simulator (LAMMPS)~\cite{10.1006/jcph.1995.1039}. 
However, it is challenging to list all the ML pre/post-processing packages in the highly active field of computational materials science.
Without any claim of completeness of the list, some of the popular ML tools are ASE (Atomistic Simulation Environment)~\cite{10.1088/1361-648X/aa680e}, Amp (Atomistic Machine-learning Package)~\cite{10.1016/j.cpc.2016.05.010}, N2P2 (Neural Network Potential Package)~\cite{10.5281/zenodo.4750573}, Aenet (Atomic Energy Network)~\cite{10.1016/j.commatsci.2015.11.047}, KLIFF (KIM-based Learning-Integrated Fitting Framework) from OpenKIM project~\cite{10.25950/ff8f563a}, MAISE (Module for ab initio structure evolution)~\cite{10.1016/j.cpc.2020.107679}. 
Packages like Aenet, MAISE, and N2P2 support Behler-Parrinello type~\cite{10.1103/PhysRevLett.98.146401} neural network (NN) potential training, whereas KLIFF package provides support for both traditional and NN potential training through OpenKIM project~\cite{10.25950/ff8f563a}. 
Amp can be used to train user-defined regression models on top of NN potential training, whereas ASE provides support on data processing and generation for atomistic simulations.
\par To map the DFT data into PES using MLIPs, one of the most common descriptors is atom-centered symmetry functions (ACSF)~\cite{10.1063/1.3553717}.
Starting from 2007, different types of descriptors have been employed to map the atomic environment, focusing on invariance (translation, permutation, rotation)~\cite{10.1021/acs.chemrev.0c00868} and recently focusing on equivarience (rotation)~\cite{10.1038/s41467-022-29939-5, geiger2022e3nn}. 
Recent advances in data flow from DFT to MLIPs focus on including long-range interactions based on local charges~\cite{10.1021/acs.chemrev.0c00868}.
Thus, we can aim to answer physical and chemical interactions using global charge equilibration in long-range charge transfer in the next-generation MLIPs. 
%
\par As MLIPs extract information from large QM datasets of forces and energies, these data-driven models do not have any “built-in” meaningful physics by construction~\cite{10.1063/5.0139611}. 
Therefore, rigorous validation of MLIPs is required, starting from estimating the “Pareto front” of efficiency by identifying the most accurate and least computational resource-hungry fitting frameworks~\cite{10.1021/acs.jpca.9b08723}. 
This should followed by “domain-specific” tests such as evaluating defect formation energies, phase stability, surface reconstructions, etc., to assess the applicability of the MLIPs~\cite{10.1063/5.0005084}.
The validation tests can be complemented through the recent best-practice guidelines for MLIP development~\cite{10.1038/s41557-021-00716-z,10.1038/s41570-022-00391-9} along with uncertainty quantification techniques~\cite{10.1088/2632-2153/ab7e1a,10.1021/acs.jcim.3c00373}. 
%

%
\par Based on the above-mentioned information, we can comprehend that MLIPs can predict better atomic interaction in materials compared to traditional IPs.
However, it is clear that generating MLIPs is a complex process requiring multiple skill sets, as it involves challenges in data quality, model transferability, and computational efficiency.
In order to understand the consensus of the materials science community on the applicability of MLIPs, a survey was conducted across different demographics.
Guided by insights gathered from a comprehensive global survey, we introduce an open-source tool named \texttt{AtomProNet}, designed to simplify the development of MLIPs. 
\texttt{AtomProNet} streamlines the data collection from existing databases such as Materials Project~\cite{10.1063/1.4812323}, high-throughput data generation using VASP~\cite{10.1016/0927-0256(96)00008-0} or Quantum Espresso~\cite{10.1063/5.0005082}, batch data processing to prepare dataset for NN training and visualizations of MLIP data. 
To demonstrate its effectiveness, we construct and evaluate two state-of-the-art MLIPs alongside two classical potentials using molecular dynamics (MD), providing a detailed comparison of their performance. 
By doing so, we offer a systematic and reproducible approach for integrating machine learning into atomic-scale simulations, promoting broader adoption of advanced computational methods across the materials science community.

\section{\label{sec:Computational_and_Theoretical Method}State-of-the-Art Machine Learning Interatomic Potentials}
\par The first building block of the state-of-the-art MLIP is constructing the database of reference values (i.e., data labels): energies, forces, and stresses from the “relevant” chemical environments' representative structures (i.e., data locations) as shown in~\fig{ML_potential_schematic}(a). MLIPs can only “see” what they are trained for, indicating the importance of careful crafting of ground-truth datasets such as ensuring the strict convergence with k-point sampling and accurate capturing of van der Waals interaction in the DFT functional~\cite{10.1021/acs.chemrev.1c00022, 10.1039/D1DD00005E}.
\par The second building block is the ML feature that defines the local atomic environment representer, commonly known as the descriptor, as shown in~\fig{ML_potential_schematic}(b). The construction of descriptors defines how well it will “learn” an atomistic structural representation~\cite{10.1021/acs.chemrev.1c00021} and how well it will mimic the atomic translational, rotational, and permutational invariances and equivariences~\cite{10.1103/PhysRevLett.125.166001}.
\par The third building block is the regression framework for fitting a highly parameterized function on the large QM dataset for deep learning model or small QM dataset for “tailored” kernel-based approach, as shown in~\fig{ML_potential_schematic}(c)~\cite{10.1039/D1SC03564A,10.1021/acs.chemrev.1c00022,10.1021/acs.chemrev.0c01111,10.1021/acs.chemrev.0c00868}. This step includes a choice of hyperparameters, learning rate, and batch size, for the selected fitting approaches from artificial neural network (NN) models (Behler–Parrinello-type NNs~\cite{10.1103/PhysRevLett.98.146401}, DeepMD~\cite{PhysRevLett.120.143001}, SchNet~\cite{10.1063/1.5019779}, NequIP~\cite{10.1038/s41467-022-29939-5}), kernel-based methods (Gaussian Approximation Potentials (GAP) framework~\cite{PhysRevLett.104.136403}) and linear models (Spectral Neighbor Analysis Potential (SNAP)~\cite{10.1016/j.jcp.2014.12.018}, Moment Tensor Potential (MTP)~\cite{10.1137/15M1054183}, Atomic Cluster Expansion (ACE)~\cite{10.1103/PhysRevB.99.014104}). The point to be noted here is that these architectures act as a “black box” as they can not carry any physical meaning for the fitting parameters other than “trial and error” based hyperparameter tuning~\cite{10.1016/j.actamat.2021.116980}. The most recent advancement in the MLIP field is the universal machine learning interatomic potentials (U-MLIP), often called foundation models (MACE~\cite{arxiv.org/abs/2401.00096}, CHGNet~\cite{10.1038/s42256-023-00716-3}, and M3GNet~\cite{10.1038/s43588-022-00349-3}), which provides a universal PES to build model for different atomic structures “on the fly”.

\subsection{\label{Analytical_Modeling}Construction of potentials}
\begin{figure}[h]  
\centering
\label{}
\includegraphics[width=1.0\linewidth]{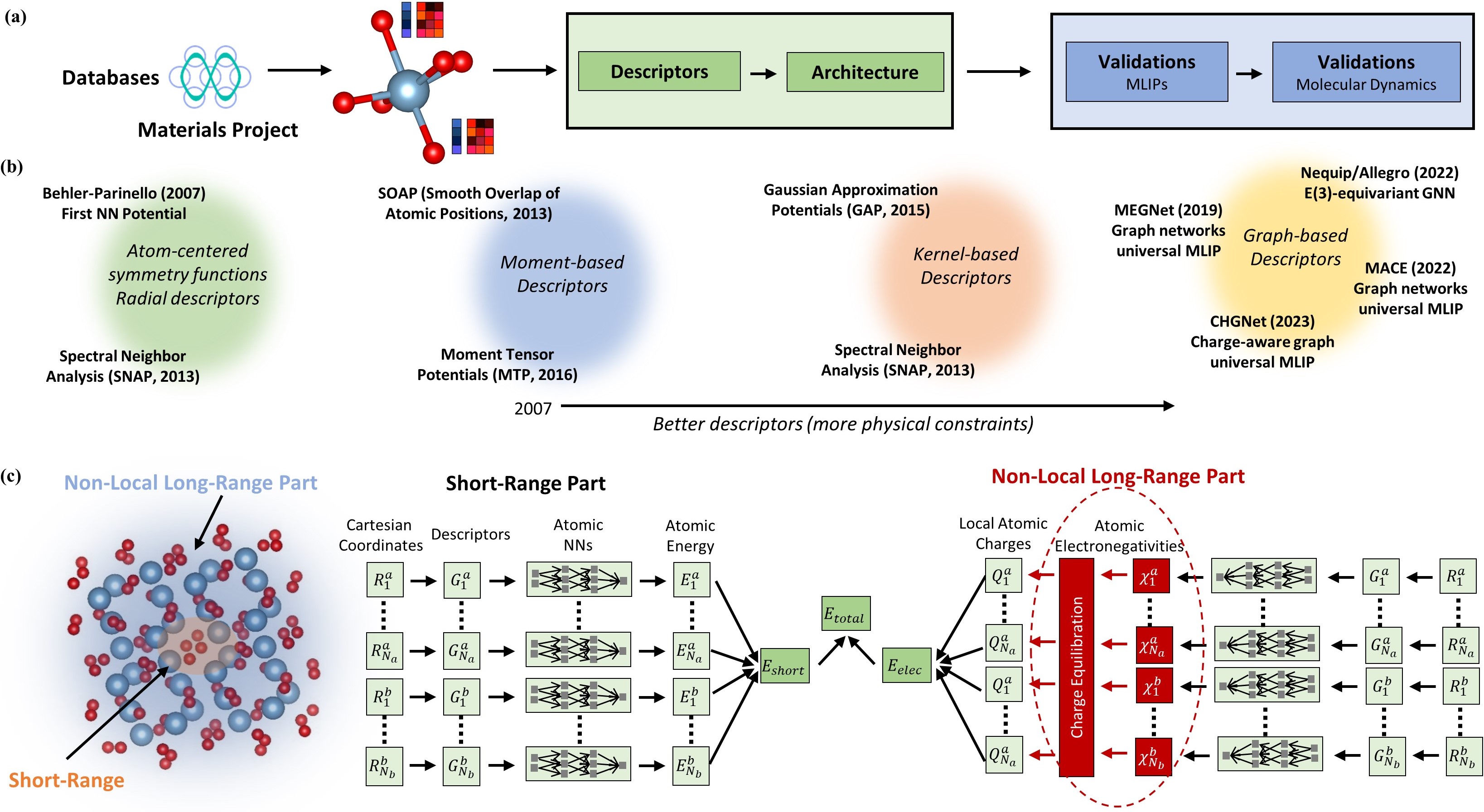}
\qquad
\caption{
\label{fig:ML_potential_schematic} (a) Schematic of dataflow from quantum mechanics (database/simulations) to development of machine learning interatomic potentials to classical molecular dynamics, (b) evolution of atomic environment descriptors, and (c) generalized neural network (NN) architecture including both short and non-local long-range atomic contributions for an atomic system of a and b elements containing $N_\textrm{a}$ and $N_\textrm{b}$ atoms, respectively.}
\end{figure}

The PES of an MLIP consists of the summation of atom ($i$)-centered local maps where each map describes the local environment for a specific cutoff sphere of a radius $r_c$ (usually varies between 3-10 Å). The local atomic position vector $\mathbf{R}_i \left(i =1,2, \ldots, N_{\textrm{a}} \textrm{\:or\:} N_{\textrm{b}}\right)$ of atom $i$ defines the atomic energies $E_i$ for the local map which will then gives the PES through “short-range energy” or $E_{\text {short}}=\sum_i E_i$. The reference QM datasets are usually prepared from DFT calculations from a set of $N$ supercells ($s=1,2,\ldots,N$) of appropriate physics by collecting the energies $\left\{E_{\mathrm{DFT}}^s\right\}$ and (often) atomic forces $\left\{\mathrm{F}_{\mathrm{DFT}}^s\right\}$ and (often) stress tensors $\left\{\mathbf{T}_{\mathrm{DFT}}^s\right\}$. 
\par \sloppy To represent the local atomic map in an ML framework, structural descriptors $\mathbf{G}_\eta\: (\eta=1,2, \ldots, K)$ are first proposed by Behler and Parrinello~\cite{10.1103/PhysRevLett.98.146401}. Here, the input layer of NN has $K$ nodes. 
$\mathbf{G}_\eta$ are generally smooth functions of  position vector $\mathbf{R}_i$ (e.g. atom-centered symmetry functions (ACSF)~\cite{10.1063/1.3553717}, smooth overlap of atomic positions (SOAP)~\cite{10.1103/PhysRevB.87.184115}, Coulomb matrix~\cite{10.1103/PhysRevLett.108.058301}), which satisfies invariance under translations and rotations of atoms. 
The advantage of using structural descriptors instead of directly using $\mathbf{R}_i$ is that they can represent any variable size $\mathbf{R}_i$ to map using its feature vectors~\cite{10.1016/j.actamat.2021.116980}.
Finally, using a regression model $\mathscr{R}$, the $\mathbf{G}_\eta$ will be mapped to $E_i$ as shown in~\eq{PES_ML}:
\begin{equation}\label{eq:PES_ML}
\begin{aligned}
\textrm{NN:\:} &\mathbf{R}_i \rightarrow \mathbf{G}_\eta \xrightarrow{\mathscr{R}} E_i,\\
\textrm{PINN:\:} &\mathbf{R}_i \rightarrow \mathbf{G}_\eta \xrightarrow{\mathscr{R}} \mathbf{p}_{\mathbf{i}} \xrightarrow{\Phi} E_i.
\end{aligned}
\end{equation}
\par To get “physically meaningful” predictions while mapping $\mathbf{G}_\eta$ onto $E_i$, a new set of potential parameters $\mathbf{p}_i$ can also be used in NN training. The role of $\mathbf{p}_i$ is to guide the flow of data from the regression model through a physics-embodied functional ($\mathbf{p}_i$) as shown in~\eq{PES_ML}. 
This idea is known as a physics-informed neural network (PINN)~\cite{10.1016/j.jcp.2018.10.045} where the interatomic potential ($\Phi\left(\mathbf{R}_i, \mathbf{p}_i\right)$) are expected to “pipe” data (extrapolate/interpolate) satisfying appropriate physics rather than being a purely mathematical technique. 
\par The typical feed-forward architecture used in MLIPs consists of $M$ layers, with input and hidden layers, having  $K$ and $k$ nodes/neurons, respectively, for a system containing the elements $(\mu =a,b)$ as shown in~\fig{ML_potential_schematic}(c). 
The data flows to the input layer ($n=1$) from $\mathbf{G}_\eta$ which are then multiplied with weights $w_{\eta \nu}^{(n-1)}\left(\nu=1,2, \ldots, k\right.$), followed a shift by a bias term ($b_\nu^{(n-1)}$). 
This data flows to the second layer ($n=2$, first hidden layer) and gets multiplied with an activation (transfer) function $f^{(2)}(x)$.
The output of the first hidden layer ($t_\nu^{(2)}=f^{(2)}\left(\sum_{\eta=1}^{\eta=K} G_\kappa w_{\eta \nu}^{(1)}+b_\nu^{(1)}\right)$) flows as input for the second hidden layer ($n=3$) and the output of the second layer ($t_\nu^{(3)}=f^{(3)}\left(\sum_{\nu=1}^{\nu=\nu} t_\nu^{(2)} w_{\nu \nu}^{(2)}+b_\nu^{(2)}\right)$) becomes the input for the third hidden layer.
The final layer ($n=M$) of the architecture delivers the output as the following: 
\begin{equation}\label{eq:Behler_Parrinello_NN_architechtire}
\begin{aligned}
\textrm{Atomic NN:}\:\: t_\nu^{(n)}&=f^{(n)}\left(\sum_\nu t_\nu^{(n-1)} w_{\nu \nu}^{(n-1)}+b_\nu^{(n-1)}\right), \quad n=2,3, \ldots, M,  \:t_\eta^{(1)}=G_\eta,
\\
E_i &= \sum_\nu t_\nu^{(M)} w_{\nu}^{(M)}+b^{(M)}.
\end{aligned}
\end{equation}
The choice of the activation function varies depending on the architecture, such as sigmoidal function $f(x)=\left(1+e^{-x}\right)^{-1}$, the hyperbolic tangent $f(x)=\tanh (x)$~\cite{10.1039/C1CP21668F} with the obvious activation function of final layer ($f^{(M)}(x) \equiv x$)~\cite{10.1016/j.actamat.2021.116980, 10.1021/acs.chemrev.0c00868}. The output layer of the atomic NN for atom $i$ contains one node giving $E_i$, which maps the PES by summing the individual predicted short-range atomic energies $E_{short}=\sum_\mu \sum_i E_i^\mu$ 
. The output layer can also contain three nodes for force ($F_x, F_y, F_z$) or stress components ($T_{xy}, T_{yz}, T_{zx}$), considering their components are non-zero (can be zero due to symmetry) during training. 
To accommodate long-range interaction while having ${r_{c}}$ imposed on local atomic structures, a second set of atomic feed-forward networks can be used as shown~\fig{ML_potential_schematic}(c)~\cite{10.1021/acs.chemrev.0c00868}. The second set will use descriptors for atomic charges ($Q_i^\mu$), which resolve for local charges and finally will be added to get long-range electrostatic energy $E_{\text {elec }}$ using  Coulomb's law (nonperiodic systems) or Ewald sum (periodic systems)~\cite{10.1002/andp.19213690304}~as shown in~\eq{energy_with_elec_eq}:
\begin{equation}\label{eq:energy_with_elec_eq}
\begin{aligned}
E_{\text {total }}=\sum_{\mu=a}^{N_{\text {elem }}} \sum_{i=1}^{N_{\text {atom }}} E_i^\mu+E_{\text {elec }}.
\end{aligned}
\end{equation}
Point to be noted here, at first, the atomic charge neural networks have to be trained from the reference DFT charges to get $E_{\text {elec}}$. 
Then $E_{\text {short}}$ can be trained after subtracting the electrostatic term ($E_{\text {elec}}$) to avoid the double counting of $E_{\text {elec}}$ in the total energy.
For example,  PhysNet~\cite{10.1021/acs.jctc.9b00181} implemented long-range electrostatics, whereas TensorMol~\cite{10.1039/C7SC04934J} even implemented long-range dispersion through Grimme D3 dispersion correction~\cite{10.1063/1.3382344}. 
The limitation here is the dependency on reference DFT partial charges, although they are mostly calculated from mathematically well-defined charge-partitioning methods such as Mulliken,~\cite{10.1063/1.1740588} or Bader charges~\cite{10.1021/ar00109a003}. 
Thus, adding non-local interactions by global charge (re)distribution in the NN architecture can accurately depict long-range phenomena (e.g., doping) instead of using fixed atomic charges with higher $\mathrm{r_c}$. 
Though global charge (re)distribution techniques are rarely used for MLIPs, classical force fields like ReaxFF~\cite{10.1021/jp004368u} are successfully using electronegativity equalization~\cite{10.1021/ja00275a013} and charge equilibration (QeQ) method proposed by Rappe and Goddard~\cite{10.1021/j100161a070}.
%

%
\par Atomic forces from MLIPs are not independent outputs of neurons but rather predicted through analytical derivatives of energy. The force component $F_{i \alpha}$ of atom $i$ in $\alpha =\{x, y, z\}$ direction can be written as a function of descriptors and coordinates using the chain rule as shown in~\eq{force_eq}:
\begin{equation}\label{eq:force_eq}
\begin{aligned}
F_{i \alpha}^{\text {short }}  =-\frac{\partial E_{\text {short }}}{\partial R_{i \alpha}}=-\sum_{\mu=a}^{N_{\text {elem}}} \sum_{i=1}^{N_{\text {atom}}} \frac{\partial E_i^\mu}{\partial R_{i \alpha}} 
 =-\sum_{\mu=a}^{N_{\text {elem}}} \sum_{i=1}^{N_{\text {atom}}} \sum_{j=1}^{N_{\text {sym}}} \frac{\partial E_i^\mu}{\partial G_{i j}^\mu} \frac{\partial G_{i j}^\mu}{\partial R_{i \alpha}}.
\end{aligned}
\end{equation}
Here, $N_{\text {sym }}$ defines the number of descriptors required to map the atomic environments of element $\mu$.
It is worth mentioning that atomic forces are influenced by atoms beyond their immediate neighbors with a cutoff radius up to $2 \times r_c$. 
This is because, while atomic energies depend only on atoms within $r_c$, the forces are indirectly impacted by the energies of all atoms within $r_c$, which themselves depend on their own local environments, effectively extending the influence to twice the $r_c$.
\par For a dataset consisting of $N$ DFT supercells with $N_s$ number of atoms, the loss function of MLIP training can be expressed as:
\begin{equation}\label{eq:loss_function}
\begin{aligned}
 \mathscr{E}=\frac{1}{N} \sum_{s=1}^N\left(\frac{E_{\mathrm{total}}^s-E_{\mathrm{DFT}}^s}{N_s}\right)^2+\tau_1 \frac{1}{N} \sum_{s=1}^N \sum_{\alpha=1}^3\left[F_\alpha^s-\left(F_\alpha^s\right)_{\mathrm{DFT}}\right]^2
 +\tau_2 \frac{1}{N} \sum_{s=1}^N \sum_{\alpha, \beta=1}^3\left[T_{\alpha \beta}^s-\left(T_{\alpha \beta}^s\right)_{\mathrm{DFT}}\right]^2.
\end{aligned}
\end{equation}
Here, $\tau_1$, and $\tau_2$, are unknown coefficients called hyper-parameters, 
and $\beta =\{x, y, z\}$. 
The three terms in~\eq{loss_function} represent the mean-squared errors of supercell energy, forces, and stresses, respectively.
%
%
Then, optimization algorithms such as backpropagation (steepest descent method)~\cite{10.1038/323533a0} are used to optimize the hyper-parameters to be a constant part of the definition of the MLIPs, just like the traditional potentials. 
Gradient-based optimizations are used to avoid trapping in a local minimum of the rugged terrain with multiple tips and valleys of the loss function. 
This Behler-Parrinello-type architecture deploys one neural network per atom as described by~\eq{Behler_Parrinello_NN_architechtire} and can be parallelized efficiently with a thumb rule of energies and forces prediction of 500 atoms per second per CPU core~\cite{10.1021/acs.chemrev.0c00868}. 
\section{\label{AtomProNet}AtomProNet: Atomic Data Processing for Neural Network} 
\begin{figure}[h]  
\centering
\label{}
\includegraphics[width=0.99\linewidth]{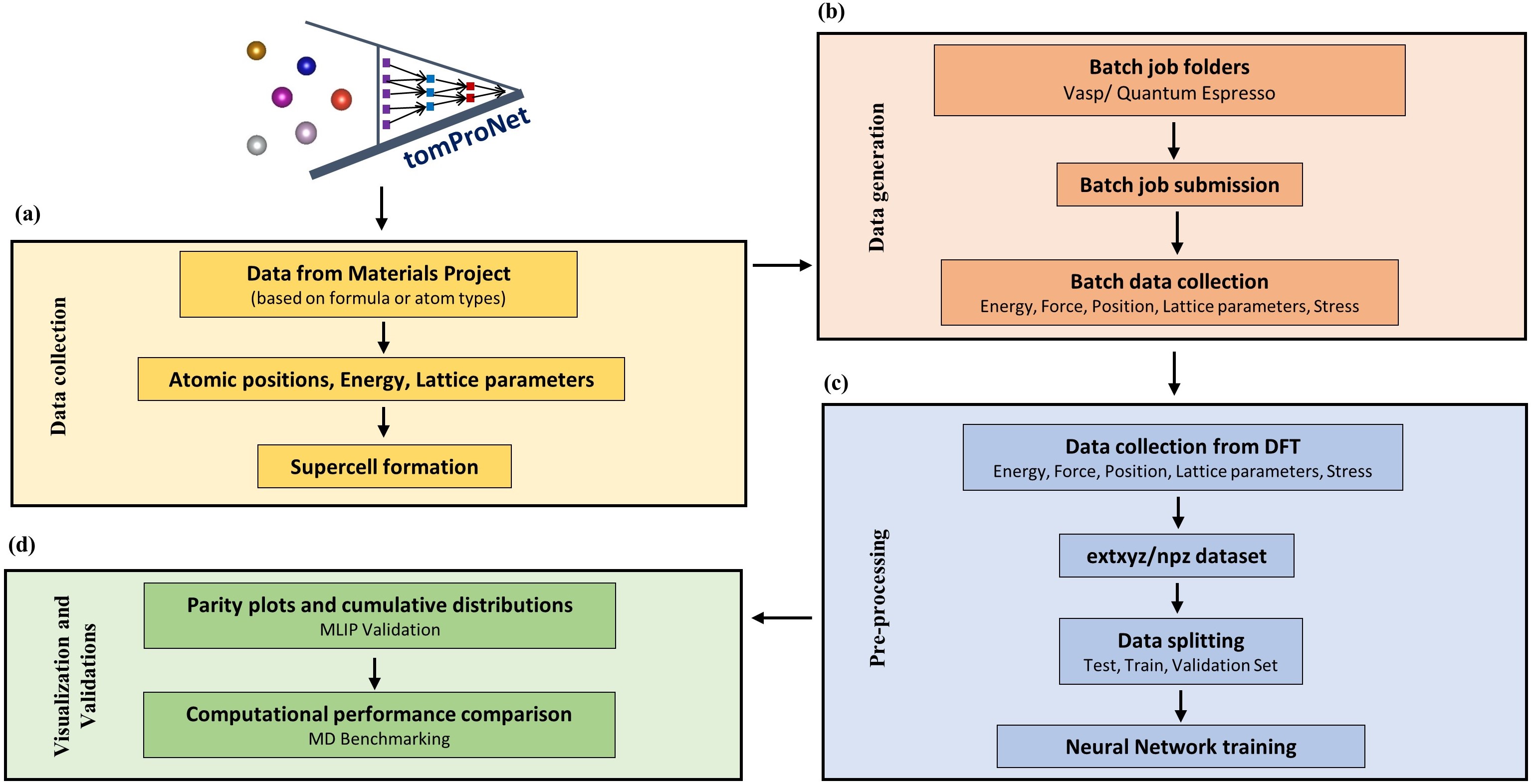}
\qquad
\caption{
\label{fig:AtomProNet_workflow} Highlights of the four modules of \texttt{AtomProNet} software - (a) data collection from materials project database, (b) data generation using DFT simulation, (c) pre-processing for neural network, (d) visualization and validations of MLIPs, and benchmarking tests of classical molecular dynamics.}
\end{figure}

In the era of artificial intelligence, materials science is rapidly advancing with the integration of machine learning. 
However, ensuring equitable access to computational tools remains a challenge, particularly for researchers who face systemic barriers, such as limited resources, lack of mentorship, or geographic isolation. 
To bridge this gap, we conducted a comprehensive global survey to assess the current state of knowledge dissemination and accessibility in computational materials science, ultimately informing the development of \texttt{AtomProNet}—a pre/post-processing software designed to empower diverse scientific communities, as shown in~\fig{AtomProNet_workflow}. 
This software includes data collection (\fig{AtomProNet_workflow}(a)), data generation (\fig{AtomProNet_workflow}(b)), batch data processing to prepare dataset for NN training (\fig{AtomProNet_workflow}(c)) and performance assessment of MLIPs (\fig{AtomProNet_workflow}(d)). 
Hence, \texttt{AtomProNet} automates the generation, validation, and benchmarking of MLIPs seamlessly. 

\subsection{\label{DataQuality}Data Collection and Generation}
%
\par Though there are no standard procedures to prepare the reference datasets from QMs (usually high-throughput DFT) for MLIPs, all “schools of thought” consensus that the dataset should adequately portray the physics of the problem domain. 
For instance, the dataset can have static structures at low-energy configurations, snapshots of AIMD trajectories even non-equilibrium structures to represent transition states. 
Usually, the datasets for the configuration space include $\sim 10^3$ to $10^4$ supercells with information about coordinates, energies, forces, (often) lattice parameters, and (often) stresses. 
Here, we used Kohn–Sham DFT~\cite{PhysRev.140.A1133,PhysRev.136.B864} to generate high-quality data to study the hydrostatic loading-induced spall failure 
as a case study to demonstrate the capabilities of \texttt{AtomProNet}. 
\par Al and O vacancy-induced pure alumina supercell of 80 atoms (using \fig{AtomProNet_workflow}(a)) were relaxed to all possible degree of freedom with Perdew, Burke, and Ernzerhof (PBE)~\cite{PhysRevLett.77.3865} semilocal exchange-correlation functionals, projector augmented-wave (PAW)~\cite{PhysRevB.59.1758} pseudopotentials using VASP~\cite{10.1016/0927-0256(96)00008-0}.
The electronic energies cut-off, the kinetic-energy cutoff, Monkhorst–Pack k-points and width of Gaussian smearing were $10^{–6}$ eV, 1 meV/atom, $\mathrm{4 \times 4 \times 4}$, and $\mathrm{0.026 eV}$, respectively~\cite{10.1021/acs.jpcc.2c06646}. 
Then, the relaxed structure is hydrostatically strained to $\pm10\%$ of the lattice parameter with a stepsize of 0.01\% (12000 cases) using the pre-processing for DFT simulation module of \texttt{AtomProNet} (\fig{AtomProNet_workflow}(b)) to perform the self-consistent field (SCF) simulations, which will be used an input data (energy and force) for MLIP training.

\subsection{MLIPs and their comparisons}
\begin{figure}[h]  
\centering
\label{}
\includegraphics[width=0.99\linewidth]{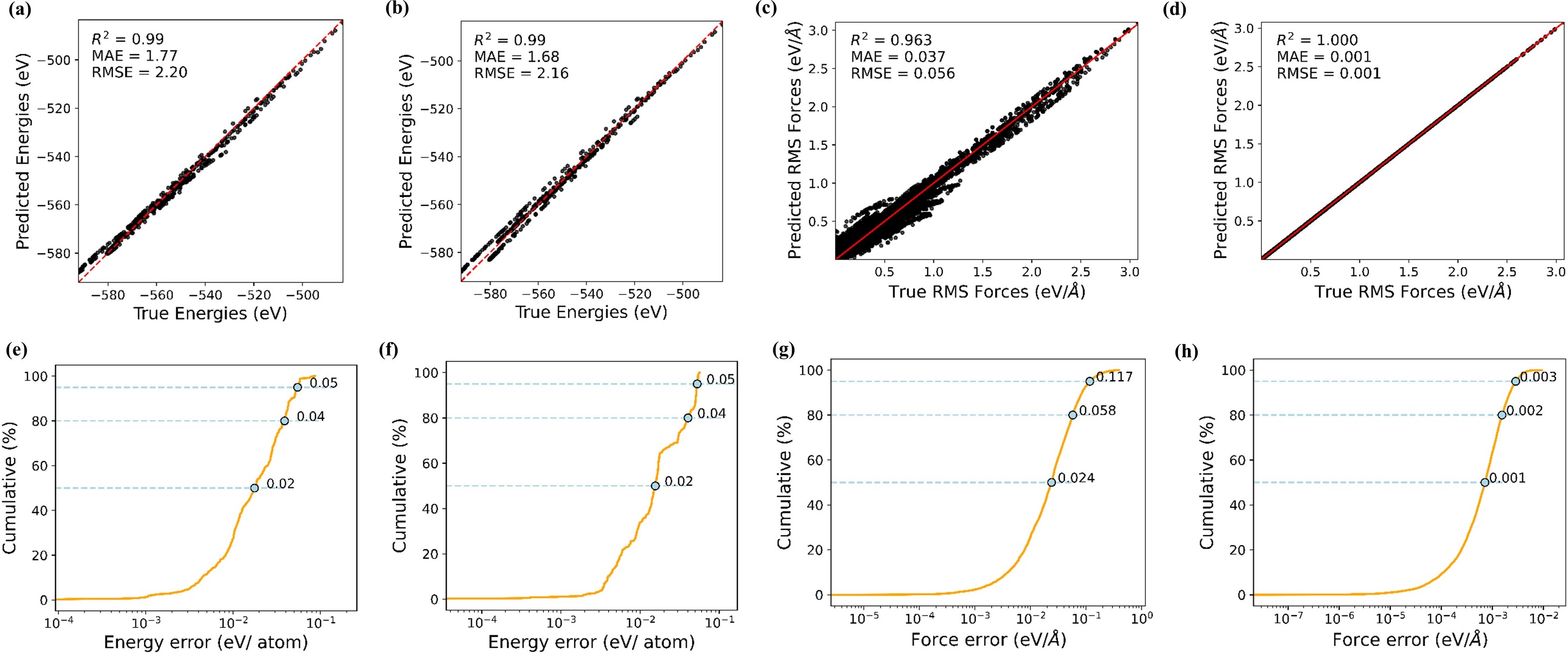}
\qquad
\caption{
\label{fig:parity_plot_mace_allegro} The parity plots illustrate the comparison between the Allegro and MACE-trained model with DFT for (a) Allegro-energy and (b) MACE-energy, (b) Allegro-RMS forces and (f) MACE-RMS forces including $\mathrm{R^2}$, MAE, RMSE values; the cumulative distribution plots show the absolute error percentages for (c) Allegro-energy and (d) MACE-energy, (g) Allegro-forces and (h) MACE-forces when comparing the MLIPs with the DFT-relaxed alumina crystal.}
\end{figure}
For building MLIPs in this study, we chose a few top-performing equivariant graph neural networks (GNN) as recommended by the Matbench discovery benchmark platform~\cite{arxiv.org/abs/2308.14920}, namely, MACE~\cite{batatia2023macehigherorderequivariant}, and Allegro~\cite{10.1038/s41467-023-36329-y}. 
Energy, force, stress, and lattice data from 16000 DFT structures are fetched and prepared into NN training format using the pre-processing for neural network module of \texttt{AtomProNet} (\fig{AtomProNet_workflow}(c)) and split into training (14000), testing (1500), and (500) validation sets. 
We used the same maximum rotation hyperparameter ($l_{max}$= 1), radial cutoff (3.0 $\mathrm{\r{A}}$) for both Allegro and MACE (version 0.3.8) with float64 precision in NVIDIA V100 single GPU training. 
We used hierarchical two-body latent multi-layer perception (MLPs) ([32, 64, 128]) in Allegro using SiLU nonlinearities~\cite{hendrycks2023gaussian} whereas MACE used 128 uniform feature channels, split into scalar (0e) and vector (1o) irreducible representations (128x0e + 128x1o)~\cite{geiger2022e3nn}. 
The Allegro loss function weights forces and total energy equally, while in MACE, the loss function initially prioritizes forces over energy (L=1×Energy Loss + 100×Forces), transitioning to energy-focused optimization in later stages (L=1000×Energy Loss + 100×Forces). 
Both models are trained using the Adam optimizer~\cite{kingma2017adam} in PyTorch~\cite{NEURIPS2019_bdbca288}.
\par After training MLIPs using both models for 150 epochs, we used post-processing of the MLIP module of \texttt{AtomProNet} (\fig{AtomProNet_workflow}(d)) to generate the parity and cumulative distribution plot based on the validation dataset of 500 structures. 
From \fig{parity_plot_mace_allegro} (a) and (b), both Allegro and MACE show similar energy prediction ($\mathrm{R^2=0.99}$) as for both models 50\%, 80\% and 90\% of the data have energy errors less than 0.02, 0.04, 0.05 eV per atom, respectively as shown in \fig{parity_plot_mace_allegro} (e) and (f). 
However, force prediction is better in MACE ($\mathrm{R^2=1}$) than Allegro ($\mathrm{R^2=0.96}$) as shown in \fig{parity_plot_mace_allegro} (c) and (d), respectively. 
This is because 90\% of the force data for MACE is below 0.003 $\mathrm{eV/{\r{A}}^2}$ whereas only 50\% of Allegro forces are below 0.024 $\mathrm{eV/{\r{A}}^2}$, as shown in \fig{parity_plot_mace_allegro} (g) and (h).
\section{\label{sec:Results_and_Discussion}Integration with Classical MD software and gold bench-marking}
Atomic trajectories, in classical MD, are calculated by solving Newton's equations of motion for each atom in the system:
\begin{equation}\label{eq:atom_traj}
    m_{i} \frac{d^{2}\mathbf{r}_{i}}{dt^{2}} = \mathbf{F}_{i} (=  -\nabla_{i} U(\mathbf{r}_{i}, \mathbf{q}_{i})),
\end{equation}
where, \textbf{F}$_{i}$ is the force acting on atom $i$, $m_{i}$, \textbf{r}$_{i}$, and \textbf{q}$_{i}$ are the mass, position, and charge of atom $i$, respectively.
The force on atom $i$ is derived from the predefined interatomic potential, i.e., calculated as the negative of the gradient of the potential energy.
Hence, at each time-step $t$, the forces are calculated using this potential, and then, through Newton's second law, the atomic positions and velocities are solved numerically using an integration scheme (i.e., Velocity-Verlet algorithm~\cite{Verlet1967}).
%
%
Thus, the choice of interatomic potential $U(\mathbf{r}_{i}, \mathbf{q}_{i})$ plays a crucial role in determining the accuracy and reliability of the MD simulation results.
An extensive review of the \textit{state-of-the-art} interatomic potentials (IP) can be found in \cite{10.1080/23746149.2022.2093129, Harrison2018}.
In MD, force fields are used to parameterize the IP's energy between atoms, which consist of mathematical functions that describe how the potential energy varies with atomic positions, bond lengths, angles, and other interaction parameters \cite{Mao2023}.
%
%
Broadly speaking, force fields can be categorized into (i) non-reactive, such as the Lennard-Jones (LJ) potential \cite{jones1924determination, LennardJones1931} and the embedded atom method (EAM) \cite{Daw1984,Baskes1992} and (ii) reactive, such as ReaxFF \cite{vanDuin2001} and charge-optimized many-body (COMB3) \cite{Yu2007, Liang2013}.
The total potential energy of a system in these IPs has been constructed with varying degrees of complexity and targets different phenomena, where they consist of inter and intra-atomic interactions:
\begin{align}
U^{\mathrm{total}}(\mathbf{r}_{i}, \mathbf{q}_{i}) &= U^{\mathrm{interatomic}}(\mathbf{r}_{i}) + U^{\mathrm{intra-atomic}}(\mathbf{r}_{i}, \mathbf{q}_{i}) + U^{\mathrm{others}}(\mathbf{r}_{i}, \mathbf{q}_{i}) \\
U^{\mathrm{interatomic}}(\mathbf{r}_{i}) &= U^{\mathrm{Coul}}(\mathbf{r}_{i},\mathbf{q}_{i}) + U^{\mathrm{vdW}}(\mathbf{r}_{i}) + U^{\mathrm{EAM}}(\mathbf{r}_{i}) \\
U^{\mathrm{intra-atomic}}(\mathbf{r}_{i}, \mathbf{q}_{i}) &= U^{\mathrm{bond}}(\mathbf{r}_{i}, \mathbf{q}_{i}) + U^{\mathrm{self}}(\mathbf{q}_{i}) + U^{\mathrm{angle}}(\mathbf{r}_{i}, \mathbf{q}_{i}) + U^{\mathrm{torsion}}(\mathbf{r}_{i}, \mathbf{q}_{i}) \\
&\quad + U^{\mathrm{conjugation}}(\mathbf{r}_{i}, \mathbf{q}_{i}) + U^{\mathrm{over}}(\mathbf{r}_{i}, \mathbf{q}_{i}) + U^{\mathrm{under}}(\mathbf{r}_{i}, \mathbf{q}_{i}).
\end{align}
Here, depending on the selected IP, certain phenomena are incorporated into Equation \ref{eq:atom_traj}.
The LJ potential considers van der Waals interactions ($U^{\mathrm{vdW}}(\mathbf{r}_{i})$), while the EAM potential considers pair and many-body interactions ($U^{\mathrm{EAM}}(\mathbf{r}_{i})$)--thus they do not explicitly include any intra-atomic interactions.
Despite this, the LJ and EAM potentials have been able to successfully study the structural and thermodynamic behavior of noble gases and metallic (pure metals and alloys) systems, respectively \cite{Cuadros1992, KATAOKA2014, Daw1993}.
ReaxFF and COMB3 incorporate both inter and intra-atomic interactions, where they both include Columbic interactions ($U^{\mathrm{Coul}}(\mathbf{r}_{i}\mathbf{q}_{i})$), $U^{\mathrm{vdW}}(\mathbf{r}_{i})$, bond-order term ($U^{\mathrm{bond}}(\mathbf{r}_{i}, \mathbf{q}_{i})$), and self-energy term ($U^{\mathrm{self}}(\mathbf{q}_{i})$).
The physics associated with
valence and angle distortion ($U^{\mathrm{angle}}(\mathbf{r}_{i}, \mathbf{q}_{i})$ and $U^{\mathrm{torsion}}(\mathbf{r}_{i}, \mathbf{q}_{i})$), three- and four-body conjugation ($U^{\mathrm{conjugation}}(\mathbf{r}_{i}, \mathbf{q}_{i})$), valence electron energy terms ($U^{\mathrm{over}}(\mathbf{r}_{i}, \mathbf{q}_{i})$ and
$U^{\mathrm{under}}(\mathbf{r}_{i}, \mathbf{q}_{i})$) in ReaxFF is embedded in the $U^{\mathrm{bond}}(\mathbf{r}_{i}, \mathbf{q}_{i})$ in COMB3.
Hence, ReaxFF and COMB3 share similarities but differ in how they handle bonding and charges \cite{Shin2012, Liang2013_1}.
ReaxFF is effective for simulating reaction processes by maintaining bond orders during bond formation and breaking in transition-state geometries.
In contrast, COMB3 explicitly couples bond order and distance relations with dynamic charges, enabling it to represent systems that exhibit metallic, ionic, and covalent bonding simultaneously.
The term $U^{\mathrm{others}}(\mathbf{r}_{i}, \mathbf{q}_{i})$ is used to correct specific energies associated with the bond angle in COMB3, while in ReaxFF, to capture properties
particular to the system of interest, such as hydrogen binding, C$_{2}$ species \cite{Shin2012, Mao2023}.
Details of the energy functions in the LJ, EAM, ReaxFF and COMB3 potentials can be found in \cite{LennardJones1931, Baskes1992, vanDuin2001,Liang2013}.
\subsection{Gold bench-marking}
Here, we showcase a comparison between two MLIPs, Allegro and MACE, and two reactive interatomic potentials, ReaxFF \cite{Hong2016} and COMB3 \cite{Choudhary2015}, using LAMMPS \cite{10.1006/jcph.1995.1039}.
The aim of this comparison is to provide an example of some important factors to consider when selecting an IP, and as such, one should also consider problem-specific factors.
For example, shock response studies require accurate predictions of Hugoniot states \cite{choi2018molecular}, while failure or defect analysis relies on capturing dislocation/defect behavior \cite{Luu2022}. 
%
%
Regardless of specific problem, however, any benchmarking needs to ensure accurate prediction of lattice parameters, elastic constants, cohesive energy, and vacancy formation energy.

In~\tab{gold_bench}, DFT-calculated properties \cite{lin2004crystal, xu2010first, ono2008first} are used as a reference to evaluate the accuracy of the IPs and identify key discrepancies.
The lattice constants (\textit{a}, \textit{b}, and \textit{c}) predicted by all IPs agree closely with DFT values, with a difference of less than 0.1 nm. 
In contrast, the elastic constants ($C_{11}$, $C_{22}$, and $C_{33}$) predicted by ReaxFF and COMB3 significantly overestimate the DFT values. 
This overestimation indicates that these reactive potentials predict a much stiffer material response.
%
%
On the other hand, Allegro and MACE matches more closely with DFT in predicting elastic constants, providing a more accurate representation of $\mathrm{Al_2O_3}$ mechanical behavior.
For cohesive energy and vacancy formation energies, Allegro and MACE, demonstrate the best agreement with DFT values, whereas ReaxFF and COMB3 consistently underestimate these properties. 
\begin{table}[h]
	\begin{center}
		\caption{Properties of $\mathrm{Al_2O_3}$ predicted by the selected interatomic potential (Allegro, MACE, ReaxFF, and COMB3) with DFT.}\label{tab:gold_bench}    
		\begin{tabular}{p{4.25cm} p{1.75cm} p{1.75cm} p{1.75cm} p{1.75cm} p{1.75cm} p{1.75cm}}
			\hline
			\hline
            Property & DFT & Allegro & MACE & ReaxFF & COMB3  \\ 
            \hline
            \hline
            \textit{a} (nm) & 4.847 & 4.885 & 4.853 & 4.850 & 4.875 \\
            \textit{b} (nm) & 4.985 & 5.025 & 4.991 & 4.981 & 5.014 \\
            \textit{c} (nm) & 7.090 & 7.146 & 7.095 & 7.076 & 7.131 \\
            $C_\mathrm{11}$ (GPa) & 454 & 471 & 327 & 650 & 853 \\
            $C_\mathrm{22}$ (GPa) & 394 & 376 & 323 & 472 & 878 \\
            $C_\mathrm{33}$ (GPa) & 466 & 473 & 328 & 565 & 776 \\
            $E_\mathrm{coh}$ (eV/atom) & -7.397 & -7.398 & -7.401 & -6.099 & -6.014 \\
            O vacancy (eV/atom) & -7.343 & -7.347 & -7.380 & -6.110 & -5.949 \\
            Al vacancy (eV/atom) & -7.296 & -7.384 & -7.311 & -6.037 & -5.950 \\
            Al-O vacancy (eV/atom) & -7.312 & -7.378 & -7.328 & -6.052 & -5.933 \\            
			\hline
			\hline
		\end{tabular}    
	\end{center}
\end{table}

\begin{figure}[h]
\centering
\begin{subfigure}[b]{0.45\textwidth}
\centering
\includegraphics[width=\linewidth]{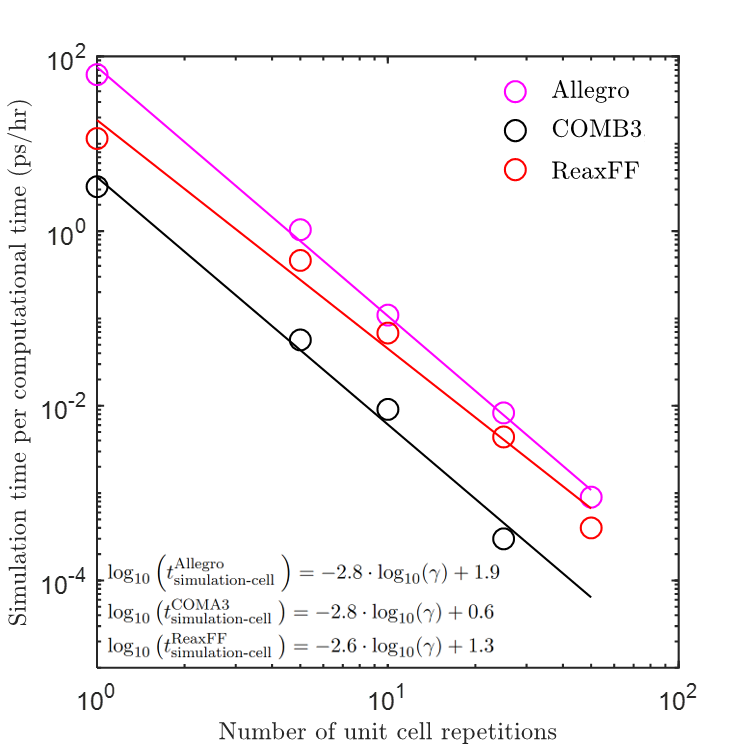}
\caption{}
\label{fig:comp_perf_sim_cell}
\end{subfigure}
\begin{subfigure}[b]{0.45\textwidth}
\centering
\includegraphics[width=\linewidth]{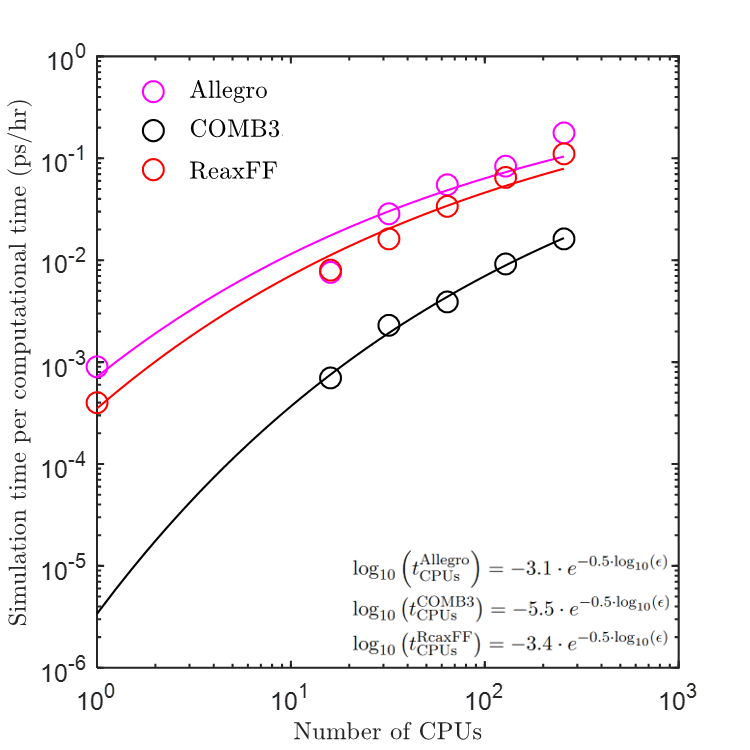}
\caption{}
\label{fig:comp_perf_cpus}
\end{subfigure}
\caption{Computational performance of Allegro, COMB3, and ReaxFF $\mathrm{Al_2O_3}$ interatomic potentials (IP) with (a) number of unit cell repetitions and (b) number of CPUs. Here, the QeQ calculation in COMB3 and ReaxFF was performed at every timestep.}
\label{fig:comp_perf}
\end{figure}

Based on~\tab{gold_bench}, it can be seen that Allegro overall managed to compare very well to the DFT calculated values, seconded by MACE.
Accurately predicting material properties is crucial, however, computational efficiency must also be considered when benchmarking interatomic potentials as achieving high accuracy without practical computational performance can limit the applicability of a potential in large-scale simulations. 
Hence, here, we conducted a computational performance comparison between the selected interatomic potentials as shown in~\fig{comp_perf}.

Computational performance is quantified as the simulation time per computational time as this metric indicates, for a given number of atoms and computational resources, the efficiency of an IP.
Prior to each simulation, the simulation cells were fully relaxed, and then an isothermal-isobaric ensemble (NPT) was applied with a timestep of 0.1 fs to quantify computational performance.
As COMB3 and ReaxFF are both compatible with QeQ charge equilibration, it was applied at every timestep to ensure an accurate representation of charge redistribution throughout the simulation.
~\fig{comp_perf_sim_cell} illustrates the relationship between the number of unit cell repetitions ($\gamma$), i.e., number of atoms, and computational performance ($t_\mathrm{simulation-cell}$).
Here, $\gamma$ = 1, 5, 10, 25, and 50 was applied resulting in a total of 20, 2500, 20000, 312500, and 2500000 atoms, respectively.
In this comparison, the computational performance is tested with respect to a single CPU processor.
The results show that Allegro exhibited the best computational performance, followed by ReaxFF, and then COMB3.
By fitting a linear trend to the data, the gradients indicate that as $\gamma \rightarrow \infty$, ReaxFF may eventually become more efficient. 
However, considering that most practical simulations are constrained to around $1\sim100$ million atoms, Allegro remains the most practical choice for a wide range of applications. 

Considering a simulation cell of $\gamma$ = 50, we next evaluated the computational performance ($t_\mathrm{CPUs}$) as a function of increasing numbers of CPUs ($\epsilon$), as illustrated in~\fig{comp_perf_cpus}.
Here, $\epsilon$ = 1, 16, 32, 64, 128, and 256 was applied.
Similarly, Allegro outperforms the reactive IPs, where here as well, as $\epsilon \rightarrow \infty$, ReaxFF may eventually become more efficient at extremely large number of CPUs.
The frequency of QeQ calculations also plays an important role in the computational performance of COMB3 and ReaxFF as shown in~\fig{comp_perf_1000_QeQ} where the QeQ calculations were performed every 1000 iterations.
It can be seen from~\fig{comp_perf_1000_QeQ} that COMB3 and ReaxFF both perform much better computationally, outperforming Allegro when $\epsilon > 256 $ showing better parallelization capabilities.
Depending on the nature of the problems being studied, the frequency of QeQ calculations in COMB3 and ReaxFF, could be reduced to improve computational performance.
%
%
\section{\label{sec:Conclusion}Conclusion}
MLIPs provide explicit functional links between atomic positions and PES, avoiding cumbersome efforts to solve the Schrödinger equation, which enables complex chemical and physical phenomena to be simulated on extended time scales. These “high-dimensional” MLIPs can capture all degrees of freedom while ensuring atom-index invariances and rotational equivariance, enabling their scalability to systems with tens of thousands of atoms. However, generating such MLIPs is a complex, multi-step challenge, as identified through our survey, which guided us in addressing the community’s needs in our open-source software. \texttt{AtomProNet}'s user-friendly modules will guide beginners and intermediate materials science researchers to use DFT-generated data to train MLIPs. Thus, researchers will be able to focus more on engineering problems rather than spending time on pre and post-processing of data flow to and from machine learning interatomic potentials.
%
%

\section{\label{sec:Supporting_Information}Supplementary information}
 QeQ calculation influence on computational performance.
\section{\label{sec:Code_Availability}Code Availability}
All pre and post-processing of VASP simulations is done using the AtomProNet python package and available at: \sloppy \href{https://github.com/MusannaGalib/AtomProNet} {https://github.com/MusannaGalib/AtomProNet}. A collab demo of how to use it is also provided along with the code.
%
%
%
%
%
\section{\label{sec:Author_Contributions}Author Contributions}
\textbf{Musanna Galib:} Conceptualization, Investigation, Software, Survey, Data Curation, Formal analysis, Validation, Writing - Original Draft, Project administration, Funding acquisition.~\textbf{Mewael Isiet:} Investigation, Formal analysis, Validation, Writing - Original Draft.~\textbf{Mauricio Ponga:} Writing - Review \& Editing, Supervision, Project administration, Funding acquisition.

\section{\label{sec:Acknowledgements}Acknowledgements}
M.G. gratefully acknowledges that the development of AtomProNet software is supported in part by funding from the Digital Research Alliance of Canada through the DRI EDIA Champions Pilot Program. We acknowledge the support from the New Frontiers in Research Fund (NFRFE-2019-01095), the Discovery grant from the Natural Sciences and Engineering Research Council of Canada (NSERC) under Award Application Number 2016-06114. This research was supported through high-performance computational resources and services provided by Advanced Research Computing at the University of British Columbia \& the Digital Research Alliance of Canada.
\bibliographystyle{ieeetr}
\bibliography{manuscript.bib}

\begin{thebibliography}{100}

\bibitem{10.1080/00268976.2017.1333644}
N.~Mardirossian and M.~Head-Gordon, ``Thirty years of density functional theory in computational chemistry: an overview and extensive assessment of 200 density functionals,'' {\em Molecular Physics}, vol.~115, no.~19, pp.~2315--2372, 2017.

\bibitem{10.1021/acs.jpca.2c06778}
D.~M. Anstine and O.~Isayev, ``Machine learning interatomic potentials and long-range physics,'' {\em The Journal of Physical Chemistry A}, vol.~127, no.~11, pp.~2417--2431, 2023.
\newblock PMID: 36802360.

\bibitem{10.1063/1.4812323}
A.~Jain, S.~P. Ong, G.~Hautier, W.~Chen, W.~D. Richards, S.~Dacek, S.~Cholia, D.~Gunter, D.~Skinner, G.~Ceder, and K.~A. Persson, ``{Commentary: The Materials Project: A materials genome approach to accelerating materials innovation},'' {\em APL Materials}, vol.~1, p.~011002, 07 2013.

\bibitem{10.1107/S0021889809016690}
S.~Gra{\v{z}}ulis, D.~Chateigner, R.~T. Downs, A.~F.~T. Yokochi, M.~Quir{\'{o}}s, L.~Lutterotti, E.~Manakova, J.~Butkus, P.~Moeck, and A.~Le~Bail, ``{Crystallography Open Database {--} an open-access collection of crystal structures},'' {\em Journal of Applied Crystallography}, vol.~42, pp.~726--729, Aug 2009.

\bibitem{10.1016/j.commatsci.2012.02.002}
S.~Curtarolo, W.~Setyawan, S.~Wang, J.~Xue, K.~Yang, R.~H. Taylor, L.~J. Nelson, G.~L. Hart, S.~Sanvito, M.~Buongiorno-Nardelli, N.~Mingo, and O.~Levy, ``Aflowlib.org: A distributed materials properties repository from high-throughput ab initio calculations,'' {\em Computational Materials Science}, vol.~58, pp.~227--235, 2012.

\bibitem{10.1557/mrs.2018.208}
C.~Draxl and M.~Scheffler, ``Nomad: The fair concept for big data-driven materials science,'' {\em MRS Bulletin}, vol.~43, pp.~676--682, Sep 2018.

\bibitem{10.1038/s41524-020-00440-1}
K.~Choudhary, K.~F. Garrity, A.~C.~E. Reid, B.~DeCost, A.~J. Biacchi, A.~R. Hight~Walker, Z.~Trautt, J.~Hattrick-Simpers, A.~G. Kusne, A.~Centrone, A.~Davydov, J.~Jiang, R.~Pachter, G.~Cheon, E.~Reed, A.~Agrawal, X.~Qian, V.~Sharma, H.~Zhuang, S.~V. Kalinin, B.~G. Sumpter, G.~Pilania, P.~Acar, S.~Mandal, K.~Haule, D.~Vanderbilt, K.~Rabe, and F.~Tavazza, ``The joint automated repository for various integrated simulations (jarvis) for data-driven materials design,'' {\em npj Computational Materials}, vol.~6, p.~173, Nov 2020.

\bibitem{10.1007/s11837-016-1998-7}
B.~Puchala, G.~Tarcea, E.~A. Marquis, M.~Hedstrom, H.~V. Jagadish, and J.~E. Allison, ``The materials commons: A collaboration platform and information repository for the global materials community,'' {\em JOM}, vol.~68, pp.~2035--2044, Aug 2016.

\bibitem{10.1016/j.mtphys.2024.101560}
J.~Schmidt, T.~F. Cerqueira, A.~H. Romero, A.~Loew, F.~Jäger, H.-C. Wang, S.~Botti, and M.~A. Marques, ``Improving machine-learning models in materials science through large datasets,'' {\em Materials Today Physics}, vol.~48, p.~101560, 2024.

\bibitem{10.1038/npjcompumats.2015.10}
S.~Kirklin, J.~E. Saal, B.~Meredig, A.~Thompson, J.~W. Doak, M.~Aykol, S.~R{\"u}hl, and C.~Wolverton, ``The open quantum materials database (oqmd): assessing the accuracy of dft formation energies,'' {\em npj Computational Materials}, vol.~1, p.~15010, Dec 2015.

\bibitem{arxiv.org/abs/2410.12771}
L.~Barroso-Luque, M.~Shuaibi, X.~Fu, B.~M. Wood, M.~Dzamba, M.~Gao, A.~Rizvi, C.~L. Zitnick, and Z.~W. Ulissi, ``Open materials 2024 (omat24) inorganic materials dataset and models,'' 2024.

\bibitem{10.1038/sdata.2018.53}
A.~Zakutayev, N.~Wunder, M.~Schwarting, J.~D. Perkins, R.~White, K.~Munch, W.~Tumas, and C.~Phillips, ``An open experimental database for exploring inorganic materials,'' {\em Scientific Data}, vol.~5, p.~180053, Apr 2018.

\bibitem{10.1080/08893110410001664882}
M.~Hellenbrandt, ``The inorganic crystal structure database (icsd)—present and future,'' {\em Crystallography Reviews}, vol.~10, no.~1, pp.~17--22, 2004.

\bibitem{10.1107/S2052520616003954}
C.~R. Groom, I.~J. Bruno, M.~P. Lightfoot, and S.~C. Ward, ``{The Cambridge Structural Database},'' {\em Acta Crystallographica Section B}, vol.~72, pp.~171--179, Apr 2016.

\bibitem{10.1038/s41467-019-13297-w}
D.~Jha, K.~Choudhary, F.~Tavazza, W.-k. Liao, A.~Choudhary, C.~Campbell, and A.~Agrawal, ``Enhancing materials property prediction by leveraging computational and experimental data using deep transfer learning,'' {\em Nature Communications}, vol.~10, p.~5316, Nov 2019.

\bibitem{10.1103/PhysRev.136.B864}
P.~Hohenberg and W.~Kohn, ``Inhomogeneous electron gas,'' {\em Phys. Rev.}, vol.~136, pp.~B864--B871, Nov 1964.

\bibitem{10.1103/PhysRev.140.A1133}
W.~Kohn and L.~J. Sham, ``Self-consistent equations including exchange and correlation effects,'' {\em Phys. Rev.}, vol.~140, pp.~A1133--A1138, Nov 1965.

\bibitem{10.1016/j.actamat.2021.116980}
Y.~Mishin, ``Machine-learning interatomic potentials for materials science,'' {\em Acta Materialia}, vol.~214, p.~116980, 2021.

\bibitem{10.1021/acs.jpca.9b08723}
Y.~Zuo, C.~Chen, X.~Li, Z.~Deng, Y.~Chen, J.~Behler, G.~Csányi, A.~V. Shapeev, A.~P. Thompson, M.~A. Wood, and S.~P. Ong, ``Performance and cost assessment of machine learning interatomic potentials,'' {\em The Journal of Physical Chemistry A}, vol.~124, no.~4, pp.~731--745, 2020.
\newblock PMID: 31916773.

\bibitem{10.1016/0927-0256(96)00008-0}
G.~Kresse and J.~Furthmüller, ``Efficiency of ab-initio total energy calculations for metals and semiconductors using a plane-wave basis set,'' {\em Computational Materials Science}, vol.~6, no.~1, pp.~15--50, 1996.

\bibitem{10.1063/5.0005082}
P.~Giannozzi, O.~Baseggio, P.~Bonfà, D.~Brunato, R.~Car, I.~Carnimeo, C.~Cavazzoni, S.~de~Gironcoli, P.~Delugas, F.~Ferrari~Ruffino, A.~Ferretti, N.~Marzari, I.~Timrov, A.~Urru, and S.~Baroni, ``{Quantum ESPRESSO toward the exascale},'' {\em The Journal of Chemical Physics}, vol.~152, p.~154105, 04 2020.

\bibitem{10.1006/jcph.1995.1039}
S.~Plimpton, ``Fast parallel algorithms for short-range molecular dynamics,'' {\em Journal of Computational Physics}, vol.~117, no.~1, pp.~1--19, 1995.

\bibitem{10.1088/1361-648X/aa680e}
A.~H. Larsen, J.~J. Mortensen, J.~Blomqvist, I.~E. Castelli, R.~Christensen, M.~Dułak, J.~Friis, M.~N. Groves, B.~Hammer, C.~Hargus, E.~D. Hermes, P.~C. Jennings, P.~B. Jensen, J.~Kermode, J.~R. Kitchin, E.~L. Kolsbjerg, J.~Kubal, K.~Kaasbjerg, S.~Lysgaard, J.~B. Maronsson, T.~Maxson, T.~Olsen, L.~Pastewka, A.~Peterson, C.~Rostgaard, J.~Schiøtz, O.~Schütt, M.~Strange, K.~S. Thygesen, T.~Vegge, L.~Vilhelmsen, M.~Walter, Z.~Zeng, and K.~W. Jacobsen, ``The atomic simulation environment—a python library for working with atoms,'' {\em Journal of Physics: Condensed Matter}, vol.~29, p.~273002, jun 2017.

\bibitem{10.1016/j.cpc.2016.05.010}
A.~Khorshidi and A.~A. Peterson, ``Amp: A modular approach to machine learning in atomistic simulations,'' {\em Computer Physics Communications}, vol.~207, pp.~310--324, 2016.

\bibitem{10.5281/zenodo.4750573}
A.~Singraber, mpbircher, S.~Reeve, D.~W. Swenson, J.~Lauret, and philippedavid, ``Compphysvienna/n2p2: Version 2.1.4,'' May 2021.

\bibitem{10.1016/j.commatsci.2015.11.047}
N.~Artrith and A.~Urban, ``An implementation of artificial neural-network potentials for atomistic materials simulations: Performance for $\mathrm{TiO_2}$,'' {\em Computational Materials Science}, vol.~114, pp.~135--150, 2016.

\bibitem{10.25950/ff8f563a}
R.~S. Elliott and E.~B. Tadmor, ``Knowledgebase of interatomic models ({KIM}) application programming interface ({API}),'' 2011.

\bibitem{10.1016/j.cpc.2020.107679}
S.~Hajinazar, A.~Thorn, E.~D. Sandoval, S.~Kharabadze, and A.~N. Kolmogorov, ``Maise: Construction of neural network interatomic models and evolutionary structure optimization,'' {\em Computer Physics Communications}, vol.~259, p.~107679, 2021.

\bibitem{10.1103/PhysRevLett.98.146401}
J.~Behler and M.~Parrinello, ``Generalized neural-network representation of high-dimensional potential-energy surfaces,'' {\em Phys. Rev. Lett.}, vol.~98, p.~146401, Apr 2007.

\bibitem{10.1063/1.3553717}
J.~Behler, ``{Atom-centered symmetry functions for constructing high-dimensional neural network potentials},'' {\em The Journal of Chemical Physics}, vol.~134, p.~074106, 02 2011.

\bibitem{10.1021/acs.chemrev.0c00868}
J.~Behler, ``Four generations of high-dimensional neural network potentials,'' {\em Chemical Reviews}, vol.~121, no.~16, pp.~10037--10072, 2021.
\newblock PMID: 33779150.

\bibitem{10.1038/s41467-022-29939-5}
S.~Batzner, A.~Musaelian, L.~Sun, M.~Geiger, J.~P. Mailoa, M.~Kornbluth, N.~Molinari, T.~E. Smidt, and B.~Kozinsky, ``E(3)-equivariant graph neural networks for data-efficient and accurate interatomic potentials,'' {\em Nature Communications}, vol.~13, p.~2453, May 2022.

\bibitem{geiger2022e3nn}
M.~Geiger and T.~Smidt, ``e3nn: Euclidean neural networks,'' 2022.

\bibitem{10.1063/5.0139611}
J.~D. Morrow, J.~L.~A. Gardner, and V.~L. Deringer, ``{How to validate machine-learned interatomic potentials},'' {\em The Journal of Chemical Physics}, vol.~158, p.~121501, 03 2023.

\bibitem{10.1063/5.0005084}
P.~Rowe, V.~L. Deringer, P.~Gasparotto, G.~Csányi, and A.~Michaelides, ``{An accurate and transferable machine learning potential for carbon},'' {\em The Journal of Chemical Physics}, vol.~153, p.~034702, 07 2020.

\bibitem{10.1038/s41557-021-00716-z}
N.~Artrith, K.~T. Butler, F.-X. Coudert, S.~Han, O.~Isayev, A.~Jain, and A.~Walsh, ``Best practices in machine learning for chemistry,'' {\em Nature Chemistry}, vol.~13, pp.~505--508, Jun 2021.

\bibitem{10.1038/s41570-022-00391-9}
A.~Bender, N.~Schneider, M.~Segler, W.~Patrick~Walters, O.~Engkvist, and T.~Rodrigues, ``Evaluation guidelines for machine learning tools in the chemical sciences,'' {\em Nature Reviews Chemistry}, vol.~6, pp.~428--442, Jun 2022.

\bibitem{10.1088/2632-2153/ab7e1a}
K.~Tran, W.~Neiswanger, J.~Yoon, Q.~Zhang, E.~Xing, and Z.~W. Ulissi, ``Methods for comparing uncertainty quantifications for material property predictions,'' {\em Machine Learning: Science and Technology}, vol.~1, p.~025006, may 2020.

\bibitem{10.1021/acs.jcim.3c00373}
E.~Heid, C.~J. McGill, F.~H. Vermeire, and W.~H. Green, ``Characterizing uncertainty in machine learning for chemistry,'' {\em Journal of Chemical Information and Modeling}, vol.~63, no.~13, pp.~4012--4029, 2023.
\newblock PMID: 37338239.

\bibitem{10.1021/acs.chemrev.1c00022}
V.~L. Deringer, A.~P. Bart{\'o}k, N.~Bernstein, D.~M. Wilkins, M.~Ceriotti, and G.~Cs{\'a}nyi, ``Gaussian process regression for materials and molecules,'' {\em Chemical Reviews}, vol.~121, pp.~10073--10141, Aug 2021.

\bibitem{10.1039/D1DD00005E}
D.~Bayerl, C.~M. Andolina, S.~Dwaraknath, and W.~A. Saidi, ``Convergence acceleration in machine learning potentials for atomistic simulations,'' {\em Digital Discovery}, vol.~1, pp.~61--69, 2022.

\bibitem{10.1021/acs.chemrev.1c00021}
F.~Musil, A.~Grisafi, A.~P. Bartók, C.~Ortner, G.~Csányi, and M.~Ceriotti, ``Physics-inspired structural representations for molecules and materials,'' {\em Chemical Reviews}, vol.~121, no.~16, pp.~9759--9815, 2021.
\newblock PMID: 34310133.

\bibitem{10.1103/PhysRevLett.125.166001}
S.~N. Pozdnyakov, M.~J. Willatt, A.~P. Bart\'ok, C.~Ortner, G.~Cs\'anyi, and M.~Ceriotti, ``Incompleteness of atomic structure representations,'' {\em Phys. Rev. Lett.}, vol.~125, p.~166001, Oct 2020.

\bibitem{10.1039/D1SC03564A}
M.~Pinheiro, F.~Ge, N.~Ferré, P.~O. Dral, and M.~Barbatti, ``Choosing the right molecular machine learning potential,'' {\em Chem. Sci.}, vol.~12, pp.~14396--14413, 2021.

\bibitem{10.1021/acs.chemrev.0c01111}
O.~T. Unke, S.~Chmiela, H.~E. Sauceda, M.~Gastegger, I.~Poltavsky, K.~T. Schütt, A.~Tkatchenko, and K.-R. Müller, ``Machine learning force fields,'' {\em Chemical Reviews}, vol.~121, no.~16, pp.~10142--10186, 2021.
\newblock PMID: 33705118.

\bibitem{PhysRevLett.120.143001}
L.~Zhang, J.~Han, H.~Wang, R.~Car, and W.~E, ``Deep potential molecular dynamics: A scalable model with the accuracy of quantum mechanics,'' {\em Phys. Rev. Lett.}, vol.~120, p.~143001, Apr 2018.

\bibitem{10.1063/1.5019779}
K.~T. Schütt, H.~E. Sauceda, P.-J. Kindermans, A.~Tkatchenko, and K.-R. Müller, ``{SchNet – A deep learning architecture for molecules and materials},'' {\em The Journal of Chemical Physics}, vol.~148, p.~241722, 03 2018.

\bibitem{PhysRevLett.104.136403}
A.~P. Bart\'ok, M.~C. Payne, R.~Kondor, and G.~Cs\'anyi, ``Gaussian approximation potentials: The accuracy of quantum mechanics, without the electrons,'' {\em Phys. Rev. Lett.}, vol.~104, p.~136403, Apr 2010.

\bibitem{10.1016/j.jcp.2014.12.018}
A.~Thompson, L.~Swiler, C.~Trott, S.~Foiles, and G.~Tucker, ``Spectral neighbor analysis method for automated generation of quantum-accurate interatomic potentials,'' {\em Journal of Computational Physics}, vol.~285, pp.~316--330, 2015.

\bibitem{10.1137/15M1054183}
A.~V. Shapeev, ``Moment tensor potentials: A class of systematically improvable interatomic potentials,'' {\em Multiscale Modeling \& Simulation}, vol.~14, no.~3, pp.~1153--1173, 2016.

\bibitem{10.1103/PhysRevB.99.014104}
R.~Drautz, ``Atomic cluster expansion for accurate and transferable interatomic potentials,'' {\em Phys. Rev. B}, vol.~99, p.~014104, Jan 2019.

\bibitem{arxiv.org/abs/2401.00096}
I.~Batatia, P.~Benner, Y.~Chiang, A.~M. Elena, D.~P. Kovács, J.~Riebesell, X.~R. Advincula, M.~Asta, M.~Avaylon, W.~J. Baldwin, F.~Berger, N.~Bernstein, A.~Bhowmik, S.~M. Blau, V.~Cărare, J.~P. Darby, S.~De, F.~D. Pia, V.~L. Deringer, R.~Elijošius, Z.~El-Machachi, F.~Falcioni, E.~Fako, A.~C. Ferrari, A.~Genreith-Schriever, J.~George, R.~E.~A. Goodall, C.~P. Grey, P.~Grigorev, S.~Han, W.~Handley, H.~H. Heenen, K.~Hermansson, C.~Holm, J.~Jaafar, S.~Hofmann, K.~S. Jakob, H.~Jung, V.~Kapil, A.~D. Kaplan, N.~Karimitari, J.~R. Kermode, N.~Kroupa, J.~Kullgren, M.~C. Kuner, D.~Kuryla, G.~Liepuoniute, J.~T. Margraf, I.-B. Magdău, A.~Michaelides, J.~H. Moore, A.~A. Naik, S.~P. Niblett, S.~W. Norwood, N.~O'Neill, C.~Ortner, K.~A. Persson, K.~Reuter, A.~S. Rosen, L.~L. Schaaf, C.~Schran, B.~X. Shi, E.~Sivonxay, T.~K. Stenczel, V.~Svahn, C.~Sutton, T.~D. Swinburne, J.~Tilly, C.~van~der Oord, E.~Varga-Umbrich, T.~Vegge, M.~Vondrák, Y.~Wang, W.~C. Witt, F.~Zills, and G.~Csányi, ``A foundation model for atomistic
  materials chemistry,'' 2024.

\bibitem{10.1038/s42256-023-00716-3}
B.~Deng, P.~Zhong, K.~Jun, J.~Riebesell, K.~Han, C.~J. Bartel, and G.~Ceder, ``Chgnet as a pretrained universal neural network potential for charge-informed atomistic modelling,'' {\em Nature Machine Intelligence}, vol.~5, pp.~1031--1041, Sep 2023.

\bibitem{10.1038/s43588-022-00349-3}
C.~Chen and S.~P. Ong, ``A universal graph deep learning interatomic potential for the periodic table,'' {\em Nature Computational Science}, vol.~2, pp.~718--728, Nov 2022.

\bibitem{10.1103/PhysRevB.87.184115}
A.~P. Bart\'ok, R.~Kondor, and G.~Cs\'anyi, ``On representing chemical environments,'' {\em Phys. Rev. B}, vol.~87, p.~184115, May 2013.

\bibitem{10.1103/PhysRevLett.108.058301}
M.~Rupp, A.~Tkatchenko, K.-R. M\"uller, and O.~A. von Lilienfeld, ``Fast and accurate modeling of molecular atomization energies with machine learning,'' {\em Phys. Rev. Lett.}, vol.~108, p.~058301, Jan 2012.

\bibitem{10.1016/j.jcp.2018.10.045}
M.~Raissi, P.~Perdikaris, and G.~Karniadakis, ``Physics-informed neural networks: A deep learning framework for solving forward and inverse problems involving nonlinear partial differential equations,'' {\em Journal of Computational Physics}, vol.~378, pp.~686--707, 2019.

\bibitem{10.1039/C1CP21668F}
J.~Behler, ``Neural network potential-energy surfaces in chemistry: a tool for large-scale simulations,'' {\em Phys. Chem. Chem. Phys.}, vol.~13, pp.~17930--17955, 2011.

\bibitem{10.1002/andp.19213690304}
P.~P. Ewald, ``Die berechnung optischer und elektrostatischer gitterpotentiale,'' {\em Annalen der Physik}, vol.~369, no.~3, pp.~253--287, 1921.

\bibitem{10.1021/acs.jctc.9b00181}
O.~T. Unke and M.~Meuwly, ``Physnet: A neural network for predicting energies, forces, dipole moments, and partial charges,'' {\em Journal of Chemical Theory and Computation}, vol.~15, no.~6, pp.~3678--3693, 2019.
\newblock PMID: 31042390.

\bibitem{10.1039/C7SC04934J}
K.~Yao, J.~E. Herr, D.~Toth, R.~Mckintyre, and J.~Parkhill, ``The tensormol-0.1 model chemistry: a neural network augmented with long-range physics,'' {\em Chem. Sci.}, vol.~9, pp.~2261--2269, 2018.

\bibitem{10.1063/1.3382344}
S.~Grimme, J.~Antony, S.~Ehrlich, and H.~Krieg, ``{A consistent and accurate ab initio parametrization of density functional dispersion correction (DFT-D) for the 94 elements H-Pu},'' {\em The Journal of Chemical Physics}, vol.~132, p.~154104, 04 2010.

\bibitem{10.1063/1.1740588}
R.~S. Mulliken, ``{Electronic Population Analysis on LCAO–MO Molecular Wave Functions. I},'' {\em The Journal of Chemical Physics}, vol.~23, pp.~1833--1840, 10 1955.

\bibitem{10.1021/ar00109a003}
R.~F.~W. Bader, ``Atoms in molecules,'' {\em Accounts of Chemical Research}, vol.~18, no.~1, pp.~9--15, 1985.

\bibitem{10.1021/jp004368u}
A.~C.~T. van Duin, S.~Dasgupta, F.~Lorant, and W.~A. Goddard, ``Reaxff: A reactive force field for hydrocarbons,'' {\em The Journal of Physical Chemistry A}, vol.~105, no.~41, pp.~9396--9409, 2001.

\bibitem{10.1021/ja00275a013}
W.~J. Mortier, S.~K. Ghosh, and S.~Shankar, ``Electronegativity-equalization method for the calculation of atomic charges in molecules,'' {\em Journal of the American Chemical Society}, vol.~108, no.~15, pp.~4315--4320, 1986.

\bibitem{10.1021/j100161a070}
A.~K. Rappe and W.~A.~I. Goddard, ``Charge equilibration for molecular dynamics simulations,'' {\em The Journal of Physical Chemistry}, vol.~95, no.~8, pp.~3358--3363, 1991.

\bibitem{10.1038/323533a0}
D.~E. Rumelhart, G.~E. Hinton, and R.~J. Williams, ``Learning representations by back-propagating errors,'' {\em Nature}, vol.~323, pp.~533--536, Oct 1986.

\bibitem{PhysRev.140.A1133}
W.~Kohn and L.~J. Sham, ``Self-consistent equations including exchange and correlation effects,'' {\em Phys. Rev.}, vol.~140, pp.~A1133--A1138, Nov 1965.

\bibitem{PhysRev.136.B864}
P.~Hohenberg and W.~Kohn, ``Inhomogeneous electron gas,'' {\em Phys. Rev.}, vol.~136, pp.~B864--B871, Nov 1964.

\bibitem{PhysRevLett.77.3865}
J.~P. Perdew, K.~Burke, and M.~Ernzerhof, ``Generalized gradient approximation made simple,'' {\em Phys. Rev. Lett.}, vol.~77, pp.~3865--3868, Oct 1996.

\bibitem{PhysRevB.59.1758}
G.~Kresse and D.~Joubert, ``From ultrasoft pseudopotentials to the projector augmented-wave method,'' {\em Phys. Rev. B}, vol.~59, pp.~1758--1775, Jan 1999.

\bibitem{10.1021/acs.jpcc.2c06646}
M.~Galib, O.~K. Orhan, and M.~Ponga, ``Engineering chemo-mechanical properties of zn surfaces via alucone coating,'' {\em The Journal of Physical Chemistry C}, vol.~127, no.~5, pp.~2481--2492, 2023.

\bibitem{arxiv.org/abs/2308.14920}
J.~Riebesell, R.~E.~A. Goodall, P.~Benner, Y.~Chiang, B.~Deng, A.~A. Lee, A.~Jain, and K.~A. Persson, ``Matbench discovery -- a framework to evaluate machine learning crystal stability predictions,'' 2024.

\bibitem{batatia2023macehigherorderequivariant}
I.~Batatia, D.~P. Kovács, G.~N.~C. Simm, C.~Ortner, and G.~Csányi, ``Mace: Higher order equivariant message passing neural networks for fast and accurate force fields,'' 2023.

\bibitem{10.1038/s41467-023-36329-y}
A.~Musaelian, S.~Batzner, A.~Johansson, L.~Sun, C.~J. Owen, M.~Kornbluth, and B.~Kozinsky, ``Learning local equivariant representations for large-scale atomistic dynamics,'' {\em Nature Communications}, vol.~14, Feb. 2023.

\bibitem{hendrycks2023gaussian}
D.~Hendrycks and K.~Gimpel, ``Gaussian error linear units (gelus),'' 2023.

\bibitem{kingma2017adam}
D.~P. Kingma and J.~Ba, ``Adam: A method for stochastic optimization,'' 2017.

\bibitem{NEURIPS2019_bdbca288}
A.~Paszke, S.~Gross, F.~Massa, A.~Lerer, J.~Bradbury, G.~Chanan, T.~Killeen, Z.~Lin, N.~Gimelshein, L.~Antiga, A.~Desmaison, A.~Kopf, E.~Yang, Z.~DeVito, M.~Raison, A.~Tejani, S.~Chilamkurthy, B.~Steiner, L.~Fang, J.~Bai, and S.~Chintala, ``Pytorch: An imperative style, high-performance deep learning library,'' in {\em Advances in Neural Information Processing Systems} (H.~Wallach, H.~Larochelle, A.~Beygelzimer, F.~d\textquotesingle Alch\'{e}-Buc, E.~Fox, and R.~Garnett, eds.), vol.~32, Curran Associates, Inc., 2019.

\bibitem{Verlet1967}
L.~Verlet, ``Computer “experiments” on classical fluids. i. thermodynamical properties of lennard-jones molecules,'' {\em Physical Review}, vol.~159, p.~98–103, July 1967.

\bibitem{10.1080/23746149.2022.2093129}
S.~V.~S. Martin H.~Müser and L.~Pastewka, ``Interatomic potentials: achievements and challenges,'' {\em Advances in Physics: X}, vol.~8, no.~1, p.~2093129, 2023.

\bibitem{Harrison2018}
J.~A. Harrison, J.~D. Schall, S.~Maskey, P.~T. Mikulski, M.~T. Knippenberg, and B.~H. Morrow, ``Review of force fields and intermolecular potentials used in atomistic computational materials research,'' {\em Applied Physics Reviews}, vol.~5, Aug. 2018.

\bibitem{Mao2023}
Q.~Mao, M.~Feng, X.~Z. Jiang, Y.~Ren, K.~H. Luo, and A.~C. van Duin, ``Classical and reactive molecular dynamics: Principles and applications in combustion and energy systems,'' {\em Progress in Energy and Combustion Science}, vol.~97, p.~101084, July 2023.

\bibitem{jones1924determination}
J.~Lennard-Jones, ``On the determination of molecular fields.—i. from the variation of the viscosity of a gas with temperature,'' {\em Proceedings of the Royal Society of London. Series A, Containing Papers of a Mathematical and Physical Character}, vol.~106, pp.~441--462, Oct. 1924.

\bibitem{LennardJones1931}
J.~Lennard-Jones, ``Cohesion,'' {\em Proceedings of the Physical Society}, vol.~43, p.~461–482, Sept. 1931.

\bibitem{Daw1984}
M.~S. Daw and M.~I. Baskes, ``Embedded-atom method: Derivation and application to impurities, surfaces, and other defects in metals,'' {\em Physical Review B}, vol.~29, p.~6443–6453, June 1984.

\bibitem{Baskes1992}
M.~I. Baskes, ``Modified embedded-atom potentials for cubic materials and impurities,'' {\em Physical Review B}, vol.~46, p.~2727–2742, Aug. 1992.

\bibitem{vanDuin2001}
A.~C.~T. van Duin, S.~Dasgupta, F.~Lorant, and W.~A. Goddard, ``Reaxff: A reactive force field for hydrocarbons,'' {\em The Journal of Physical Chemistry A}, vol.~105, p.~9396–9409, Sept. 2001.

\bibitem{Yu2007}
J.~Yu, S.~B. Sinnott, and S.~R. Phillpot, ``Charge optimized many-body potential for the for the si/sio$_{2}$ system,'' {\em Physical Review B}, vol.~75, Feb. 2007.

\bibitem{Liang2013}
T.~Liang, T.-R. Shan, Y.-T. Cheng, B.~D. Devine, M.~Noordhoek, Y.~Li, Z.~Lu, S.~R. Phillpot, and S.~B. Sinnott, ``Classical atomistic simulations of surfaces and heterogeneous interfaces with the charge-optimized many body (comb) potentials,'' {\em Materials Science and Engineering: R: Reports}, vol.~74, p.~255–279, Sept. 2013.

\bibitem{Cuadros1992}
F.~Cuadros and A.~Mulero, ``The radial distribution function for two-dimensional lennard-jones fluids: Computer simulation results,'' {\em Chemical Physics}, vol.~159, p.~89–97, Jan. 1992.

\bibitem{KATAOKA2014}
Y.~KATAOKA and Y.~YAMADA, ``Phase diagram of a lennard-jones system by molecular dynamics simulations,'' {\em Journal of Computer Chemistry, Japan}, vol.~13, no.~2, p.~115–123, 2014.

\bibitem{Daw1993}
M.~S. Daw, S.~M. Foiles, and M.~I. Baskes, ``The embedded-atom method: a review of theory and applications,'' {\em Materials Science Reports}, vol.~9, p.~251–310, Mar. 1993.

\bibitem{Shin2012}
Y.~K. Shin, T.-R. Shan, T.~Liang, M.~J. Noordhoek, S.~B. Sinnott, A.~C. van Duin, and S.~R. Phillpot, ``Variable charge many-body interatomic potentials,'' {\em MRS Bulletin}, vol.~37, p.~504–512, May 2012.

\bibitem{Liang2013_1}
T.~Liang, Y.~K. Shin, Y.-T. Cheng, D.~E. Yilmaz, K.~G. Vishnu, O.~Verners, C.~Zou, S.~R. Phillpot, S.~B. Sinnott, and A.~C. van Duin, ``Reactive potentials for advanced atomistic simulations,'' {\em Annual Review of Materials Research}, vol.~43, p.~109–129, July 2013.

\bibitem{Hong2016}
S.~Hong and A.~C.~T. van Duin, ``Atomistic-scale analysis of carbon coating and its effect on the oxidation of aluminum nanoparticles by reaxff-molecular dynamics simulations,'' {\em The Journal of Physical Chemistry C}, vol.~120, p.~9464–9474, Apr. 2016.

\bibitem{Choudhary2015}
K.~Choudhary, T.~Liang, A.~Chernatynskiy, S.~R. Phillpot, and S.~B. Sinnott, ``Charge optimized many-body ({COMB}) potential for al2o3materials, interfaces, and nanostructures,'' {\em Journal of Physics: Condensed Matter}, vol.~27, p.~305004, July 2015.

\bibitem{choi2018molecular}
J.~Choi, S.~Yoo, S.~Song, J.~Park, and K.~Kang, ``Molecular dynamics study of hugoniot relation in shocked nickel single crystal,'' {\em Journal of Mechanical Science and Technology}, vol.~32, no.~7, pp.~3273--3281, 2018.

\bibitem{Luu2022}
H.~Luu, S.~Raumel, F.~Dencker, M.~Wurz, and N.~Merkert, ``Nanoindentation in alumina coated al: Molecular dynamics simulations and experiments,'' {\em Surface and Coatings Technology}, vol.~437, p.~128342, May 2022.

\bibitem{lin2004crystal}
J.~Lin, O.~Degtyareva, C.~Prewitt, P.~Dera, N.~Sata, E.~Gregoryanz, H.~Mao, and R.~J. Hemley, ``Crystal structure of a high-pressure/high-temperature phase of alumina by in situ x-ray diffraction,'' {\em Nature Materials}, vol.~3, no.~6, pp.~389--393, 2004.

\bibitem{xu2010first}
B.~Xu, H.~Stokes, and J.~Dong, ``First-principles calculation of kinetic barriers and metastability for the corundum-to-rh2o3 (ii) transition in al2o3,'' {\em Journal of Physics: Condensed Matter}, vol.~22, no.~31, p.~315403, 2010.

\bibitem{ono2008first}
S.~Ono, J.~P. Brodholt, and G.~D. Price, ``First-principles simulation of high-pressure polymorphs in mgal 2 o 4,'' {\em Physics and Chemistry of Minerals}, vol.~35, pp.~381--386, 2008.

\end{thebibliography}
%

\clearpage
\raggedbottom
\setcounter{section}{0}
\setcounter{equation}{0}
\setcounter{figure}{0}
\setcounter{table}{0}
\setcounter{page}{1}
\makeatletter
\renewcommand{\thesection}{S\arabic{section}}
\renewcommand{\theequation}{S\arabic{equation}}
\renewcommand{\thefigure}{S\arabic{figure}}
\renewcommand{\thetable}{S\arabic{table}}
\renewcommand{\thepage}{SM\arabic{page}}

\renewcommand{\bibnumfmt}[1]{[S#1]}
\renewcommand{\citenumfont}[1]{S#1}

\begin{center}
{\Large \bf Supporting Information}
\end{center}

\section{QeQ calculation influence on computational performance}
\begin{figure}[h]
\centering
\begin{subfigure}[b]{0.45\textwidth}
\centering
\includegraphics[width=\linewidth]{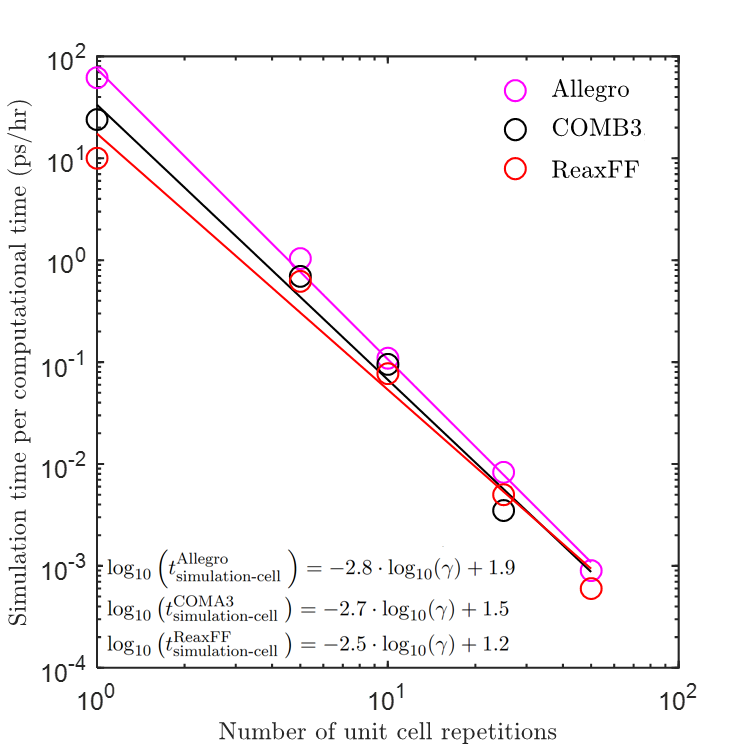}
\caption{}
\label{fig:comp_perf_sim_cell}
\end{subfigure}
\begin{subfigure}[b]{0.45\textwidth}
\centering
\includegraphics[width=\linewidth]{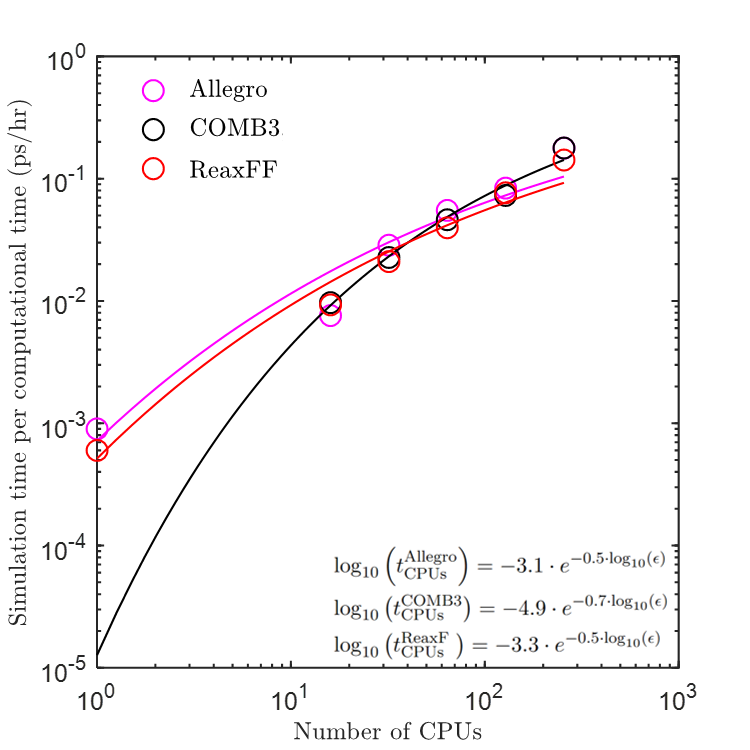}
\caption{}
\label{fig:comp_perf_1000_QeQ}
\end{subfigure}
\caption{Computational performance of Allegro, COMB3, and ReaxFF $\mathrm{Al_2O_3}$ interatomic potentials with (a) number of unit cell repetitions and (b) number of CPUs. Here, the QeQ calculation in COMB3 and ReaxFF was performed every 1000 iterations.}
\label{fig:comp_perf_1000_QeQ}
\end{figure}
%

\end{document}